# Impact of Nonlinear Absorption on Energy Distribution During FLASH-Like Deposition

Bo-Rong Lin
Independent Researcher, Penghu, Taiwan
horatio3424@outlook.com

## Abstract

FLASH radiotherapy for tumors has proven highly effective in preserving normal tissue; however, the physical origin of its biological selectivity has yet to be elucidated. This study developed a minimal diffusion-absorption model to examine the effects of nonlinear energy absorption on the spatial distribution of energy during FLASH-like deposition. Simulations revealed that nonlinear absorption alters the transfer of energy through the system while reducing the amount of energy absorbed in surrounding, unexposed regions. This behavior qualitatively resembles the tissue-sparing effect observed in FLASH radiotherapy. Parameter-space analysis revealed that the effect of incident pulse intensity depends on the absorption properties of the material, including its saturation behavior and energy threshold. These results suggest that energy-deposition conditions should be matched to the absorption characteristics of the target material.

## I. Introduction

FLASH radiotherapy is a promising cancer treatment that can achieve effective tumor control with reduced damage to normal tissues[1-3]. Various aspects of FLASH radiotherapy such as suitable delivery systems[4-6], dosimetry[7] and underlying mechanisms are being diligently studied. Several mechanisms have been proposed to explain this tissue-sparing effect, including transient oxygen depletion[8,9], reduced reactive oxygen species[10,11], radical-radical recombination[12], differences in DNA damage[13], and immune modulation[14,15]. Nonetheless, the underlying physical cause has yet to be determined conclusively, despite the fact that all of these biological responses begin with the deposition and absorption of energy. Understanding the spatial distribution of this energy is therefore an important first step in elucidating the FLASH effect. The spatial distribution of absorbed energy depends on how the target material responds to the incoming pulse, which may involve both linear and nonlinear absorption processes[16]. To examine these effects with minimal reliance on complex empirical assumptions, we considered four representative models of the absorption coefficient: constant, linear, sublinear, and threshold-dependent. These models were used to determine how the relationship between pulse intensity and energy absorption affects the spatial distribution of energy. The total energy absorbed within defined spatial regions was then calculated to compare the various absorption responses. This analysis provides a physical basis for understanding why different target materials respond differently to the same energy-deposition conditions.

## II. Material and methods

The proposed model does not attempt to reproduce the complexity of biological tissue or specific radiochemical mechanisms. Instead, the computational domain is treated as a continuous energy-absorbing material in order to focus our analysis on the physical effects of energy diffusion and absorption. This study considered a homogeneous two-dimensional material with a computational domain (Ω) measuring 250×250 pixels. As shown in Fig. 1, we defined two regions based on the radial distance from the domain center. Using spatial vector **r** = (x,y), the central source region, $\Omega_1$, was defined as a circular area measuring $|\mathbf{r}| \leq 70$ pixels, surrounded by a concentric annular region, $\Omega_2$, measuring $70 < |\mathbf{r}| \leq 100$ pixels. The dynamics of the local energy $E(\boldsymbol{r},t)$ were described using a diffusion-absorption-based partial differential equation[17,18]:

$$\frac{\partial E(\boldsymbol{r},t)}{\partial t} = \nabla \cdot (D\nabla E) - A(E) + S(\boldsymbol{r},t), \tag{1}$$

where $D$ is the energy diffusion coefficient, which was constant across the entire material, despite the heterogeneity of real tissues. Maintaining a constant $D$ ensured that any observed spatial effects were driven entirely by the energy source and nonlinear absorption mechanics. In the above equation, $S(\boldsymbol{r}, t)$ acts as the external energy source term applied only to the central region $\Omega_1$. The energy absorption term is given by

$$A(E) = \alpha(E) \cdot E, \tag{2}$$

where the dynamic absorption coefficient $\alpha(E)$ consists of a background term $\alpha_0$ and an energy-dependent nonlinear term. We tested four phenomenological models for the dynamic absorption coefficient $\alpha(E)$, including constant, linear, saturating, and threshold-dependent behavior. These models were not intended to represent specific biological pathways. The parameter values used in the simulations, expressed in arbitrary units, were not derived from empirical measurements. Instead, they were selected to ensure numerical stability and clearly illustrate the qualitative behavior of the system.

The Constant model assumes that the absorption coefficient is independent of energy, as given by:

$$\alpha(E) = \alpha_0. \tag{3}$$

The Linear model assumes that the absorption coefficient increases with energy according to energy-sensitivity coefficient $\gamma$:

$$\alpha(E) = \alpha_0 + \alpha_{\max}\gamma E, \tag{4}$$

where $\alpha_{\max}$ is the maximum additional absorption.

The sublinear Hill model[19] was used to represent a saturating absorption response as the energy approaches the characteristic threshold energy $E_{th}$:

$$\alpha(E) = \alpha_0 + \alpha_{\max}\frac{E^{0.5}}{E_{\mathrm{th}}^{0.5} + E^{0.5}}. \tag{5}$$

The Sigmoid model[20] was used to simulate a smooth and continuous threshold-dependent absorption coefficient:

$$\alpha(E) = \alpha_0 + \alpha_{\max}\frac{1}{1 + e^{-\beta(E - E_{\mathrm{th}})}}, \tag{6}$$

where β is the steepness of the curve and $E_{\mathrm{th}}$ indicates the characteristic threshold energy.

We solved the partial differential equation using a standard explicit finite-difference method. Spatial derivatives were approximated using a five-point central-difference scheme on a uniform spatial grid where $\Delta x = \Delta y = 1$ a.u. with periodic boundary conditions applied at the edges of Ω. For time stepping, we used the explicit forward Euler method with $\Delta t = 0.05$, for a total of $N_{\mathrm{step}} = 8000$ steps (representing total simulation time $T = \Delta t \cdot N_{\mathrm{step}}$). The selected time step strictly satisfied the CFL stability criterion[21] for our two-dimensional diffusion-absorption system. Running the simulation up to 8000 steps provided sufficient time for the system to approach a steady state, while minimizing residual energy at the end of the simulation. To maintain physical validity, the value of $E$ was explicitly constrained to remain non-negative. As shown in Fig. 2, we designed two source profiles to observe how the system responds to different energy deposition schemes. Both schemes delivered the same total incident energy over the simulation period to allow a direct comparison. The first scheme (CONV-like) applied a continuous (0 to 3000 steps, $T_{\mathrm{CONV\text{-}like}} = 3000 \cdot \Delta t$) low-intensity ($S_{\mathrm{CONV\text{-}like}} = 10$ a.u.) energy source. The second scheme (FLASH-like) delivered a rapid (0 to 200 steps, $T_{\mathrm{FLASH\text{-}like}} = 200 \cdot \Delta t$) high-intensity ($S_{\mathrm{FLASH\text{-}like}} = 150$ a.u.) energy burst. These source profiles made it possible to examine the effects of nonlinear absorption under different energy-deposition conditions. The absolute absorbed energy $E_{\mathrm{abs}}(\mathbf{r})$ at each spatial location was calculated by integrating the absorption over the total simulation time $T$:

$$E_{\mathrm{abs}}(\boldsymbol{r}) = \int_0^T A\big(E(\boldsymbol{r}, t)\big)dt. \tag{7}$$

Energy absorption under the various energy-deposition schemes was assessed by calculating the total absorbed energy within each region $\Omega_i$, which involved integrating $E_{\mathrm{abs}}(\mathbf{r})$ over the area:

$$E_{\mathrm{total},\Omega_i} = \iint_{\Omega_i} E_{\mathrm{abs}}(\boldsymbol{r})d^2\boldsymbol{r}. \tag{8}$$

To assess the difference in total energy burden on the surrounding region $\Omega_2$, we calculated the relative energy reduction (RER). This value, expressed as a percentage, compares the total absorbed energy under the FLASH-like scheme with that under the CONV-like scheme and is defined as follows:

$$\mathrm{RER} = \left(1 - \frac{E_{\mathrm{total},\Omega_2}^{\mathrm{FLASH\text{-}like}}}{E_{\mathrm{total},\Omega_2}^{\mathrm{CONV\text{-}like}}}\right) \times 100\%, \tag{9}$$

where $E_{\mathrm{total},\Omega_2}^{\mathrm{FLASH\text{-}like}}$ and $E_{\mathrm{total},\Omega_2}^{\mathrm{CONV\text{-}like}}$ represent the total absorbed energy within $\Omega_2$ under the respective energy deposition schemes. Accordingly, a higher RER value indicates a more pronounced suppression of the energy burden on $\Omega_2$. The simulation code was implemented on the Python 3 platform, and the primary parameters are summarized in Table 1.

**Table 1**

| **Parameter** | **Description** | **Active Domain** | **Value** |
|---|---|---|---|
| $D$ | Energy diffusion coefficient | Ω | 1 (a.u.) |
| $\alpha_0$ | Background absorption | Ω | 0.05 (a.u.) |
| $\alpha_{max}$ | Maximum additional absorption | Ω | 0.05 (a.u.) |
| β | Transition steepness | Ω | 0.05 (a.u.)$^{-1}$ |
| γ | Energy sensitivity coefficient | Ω | 0.003 (a.u.)$^{-1}$ |
| $E_{th}$ | Characteristic threshold energy | Ω | 250 (a.u.) |

## III. Results

The numerical implementation was first assessed using the Constant model ($\alpha(E) = \alpha_0$). Fig. 3 illustrates the calculated total absorbed energies in regions $\Omega_1$ (a) and $\Omega_2$ (b) versus the time step. As expected for a linear energy absorption term (i.e., $A(E) = \alpha_0 \cdot E$ ), the CONV-like and FLASH-like schemes produced identical total energy absorption in each region. Because the absorption coefficient was independent of the local energy, the temporal profile of the source did not affect the final spatial energy distribution when the total incident energy was the same. Thus, any observed differences in the subsequent simulations can be attributed to the nonlinear absorption models rather than numerical artifacts.

Fig. 4 shows the calculated total time-dependent energy absorption in regions $\Omega_1$ and $\Omega_2$ under the selected absorption models. In region $\Omega_1$ (Figs. 4a–4c), the total absorbed energies of the FLASH-like scheme exceeded those of CONV-like scheme by roughly 1.5%. In the surrounding region $\Omega_2$ (Figs. 4d–4f), the FLASH-like scheme consistently resulted in lower total absorbed energy. This reduction was far more pronounced under the Linear and Sigmoid models than under the sublinear Hill model ($n = 0.5$). Further analysis showed that the additional energy absorbed in $\Omega_1$ balanced the energy suppressed in $\Omega_2$.

Despite these model-dependent differences, the reduced energy absorption in $\Omega_2$ qualitatively resembles the "sparing effect" observed in FLASH radiotherapy. The redistribution of energy may also help to explain why the FLASH sparing effect is not consistently observed in *in vitro* experiments[22,23], where there may be less surrounding material available for energy diffusion and redistribution. Our minimal model therefore suggests that the energy-absorption properties of the target material may contribute to the observed response, independently of specific biological mechanisms.

If nonlinear energy absorption indeed contributes to this effect, then this raises questions pertaining to minimizing the total absorbed energy in $\Omega_2$. To investigate this question and guide future experimental validation, we systematically mapped the relevant parameter space using FLASH-like intensity ($S_{\text{FLASH-like}}$) as the primary control variable in conjunction with the energy sensitivity coefficient $\gamma$ for the Linear model using the characteristic threshold energy $E_{\text{th}}$ for the Hill and Sigmoid models.

Fig. 5 shows the Relative Energy Reduction (RER) across the parameter space under fixed excitation durations. When $T_{\text{FLASH-like}}$ and $T_{\text{CONV-like}}$ were maintained at previously defined values, the total incident energy deposition scaled proportionally with FLASH-like intensity ($S_{\text{FLASH-like}}$). $S_{\text{FLASH-like}}$ varied from 10 to 500 a.u., with CONV-like intensity ($S_{\text{CONV-like}}$) adjusted to maintain identical total incident energy deposition between the two schemes at each intensity level. The energy reduction produced by the sublinear Hill (n=0.5) model [Fig. 5(a)] was less pronounced, yielding an RER close to 5% across the entire landscape. The early saturation of this function at low energy levels severely limited the differential impact of intensity variation. Conversely, the unbounded Linear model [Fig. 5(c)] showed a monotonic increase in RER (>40%) with both $S_{\text{FLASH-like}}$ and the energy sensitivity coefficient $\gamma$. However, the lack of a saturation limit indicates that this was an idealized mathematical response, as real materials are expected to exhibit saturation or other nonlinear behavior at sufficiently high energy intensities.

As shown in Fig. 5(b), the Sigmoid model exhibited a more complex response, as indicated by the cyan line marking the zero-RER contour. The effect of FLASH-like source intensity ($S_{\text{FLASH-like}}$) varied with characteristic threshold energy $E_{th}$. At low $E_{\text{th}}$, increasing the source intensity could reduce the RER below zero, indicating greater energy absorption in $\Omega_2$. Conversely, at high $E_{th}$, increasing the source intensity generally increased the RER and reduced energy deposition in $\Omega_2$. In both cases, however, the RER eventually reached a limit rather than increasing indefinitely with source intensity.

Instead, it is likely that there is an optimal FLASH-like source intensity ($S_{\text{FLASH-like}}$) that maximizes the RER for a given $E_{\text{th}}$. As shown in Fig. 5(b), under a fixed excitation duration, simply increasing the source intensity may be counterproductive at low $E_{\text{th}}$ and may provide diminishing benefits at high $E_{\text{th}}$. This observation is qualitatively consistent with recent *in vivo* experimental report[24] indicating that the FLASH sparing effect diminishes or vanishes entirely when the total deposited energy exceeds a critical threshold. Further experimental studies are required to identify the optimal intensity range for different absorption characteristics. Note that the Hill model with n = 4 showed behavior similar to that of the Sigmoid model (data not shown).

Fig. 6 presents the RER across the parameter space under a fixed total incident energy. $S_{\text{FLASH-like}}$ was varied from 10 to 500 a.u., while the FLASH-like source duration was adjusted inversely to maintain the same total incident energy, while $S_{\text{CONV-like}}$ and $T_{\text{CONV-like}}$ were held at their previously defined values. Under the sublinear Hill (n = 0.5) and Linear models, the overall trends in Figs. 6(a) and 6(c) were similar to those observed in Figs. 5(a) and 5(c), respectively, and are therefore not discussed further.

As shown in Fig. 6(b), the response of the Sigmoid model varied with $E_{\text{th}}$. When $E_{\text{th}}$ was below approximately 50, increasing the FLASH-like source intensity consistently produced a negative RER, indicating greater energy absorption in $\Omega_2$. Thus, under a fixed total incident energy, increasing the source intensity did not reduce the energy burden on $\Omega_2$ at low $E_{\text{th}}$. When $E_{\text{th}}$ exceeded approximately 50, sufficiently high source intensities produced a positive RER. Beyond this point, however, the RER was determined mainly by $E_{\text{th}}$ with negligible sensitivity to further increases in source intensity. These results suggest that at higher $E_{th}$, exceeding a certain source intensity may be sufficient to reduce energy deposition in $\Omega_2$, whereas further increases provide little additional benefit.

As noted above, the phenomenological model proposed in this study does not address complex biological mechanisms; however, the observed physical behavior may apply more broadly to other systems with nonlinear absorption responses. It may even account for many of the differences observed across tumor models. In the proposed model, heterogeneity among real tumors can be represented as variation in the characteristic threshold energy ($E_{th}$). Accordingly, the energy absorbed under a given FLASH deposition scheme may differ according to whether the target is characterized by a high effective $E_{th}$, a low $E_{th}$, or a combination of both. This provides a feasible physical basis to explain observed differences in the response to a given energy-deposition scheme.

These findings also suggest that simply maximizing the incident intensity may not always enhance the sparing effect. Reducing energy deposition in the surrounding region may instead require the matching of incident intensity to the absorption characteristics of the target material. Our results further show that nonlinear energy absorption alone can produce sparing-like behavior even in a homogeneous material. Although real tissues are heterogeneous, preliminary simulations using different parameter values for the central and surrounding regions produced similar overall trends (data not shown).

## IV. Conclusion

This study developed a minimal diffusion-absorption model demonstrating that nonlinear energy absorption alone can redistribute deposited energy and reduce energy absorption in surrounding regions, producing a pattern that qualitatively resembles the FLASH sparing effect. RER heatmap analysis showed that increasing pulse intensity does not always increase energy suppression in surrounding regions. Instead, the extent of energy reduction depends on the nonlinear absorption characteristics of the material. These findings suggest that energy-deposition conditions should be adjusted according to the absorption properties of the target material. Future studies will incorporate additional nonlinear absorption models and regional variations in model parameters to better represent the heterogeneity of biological tissue.

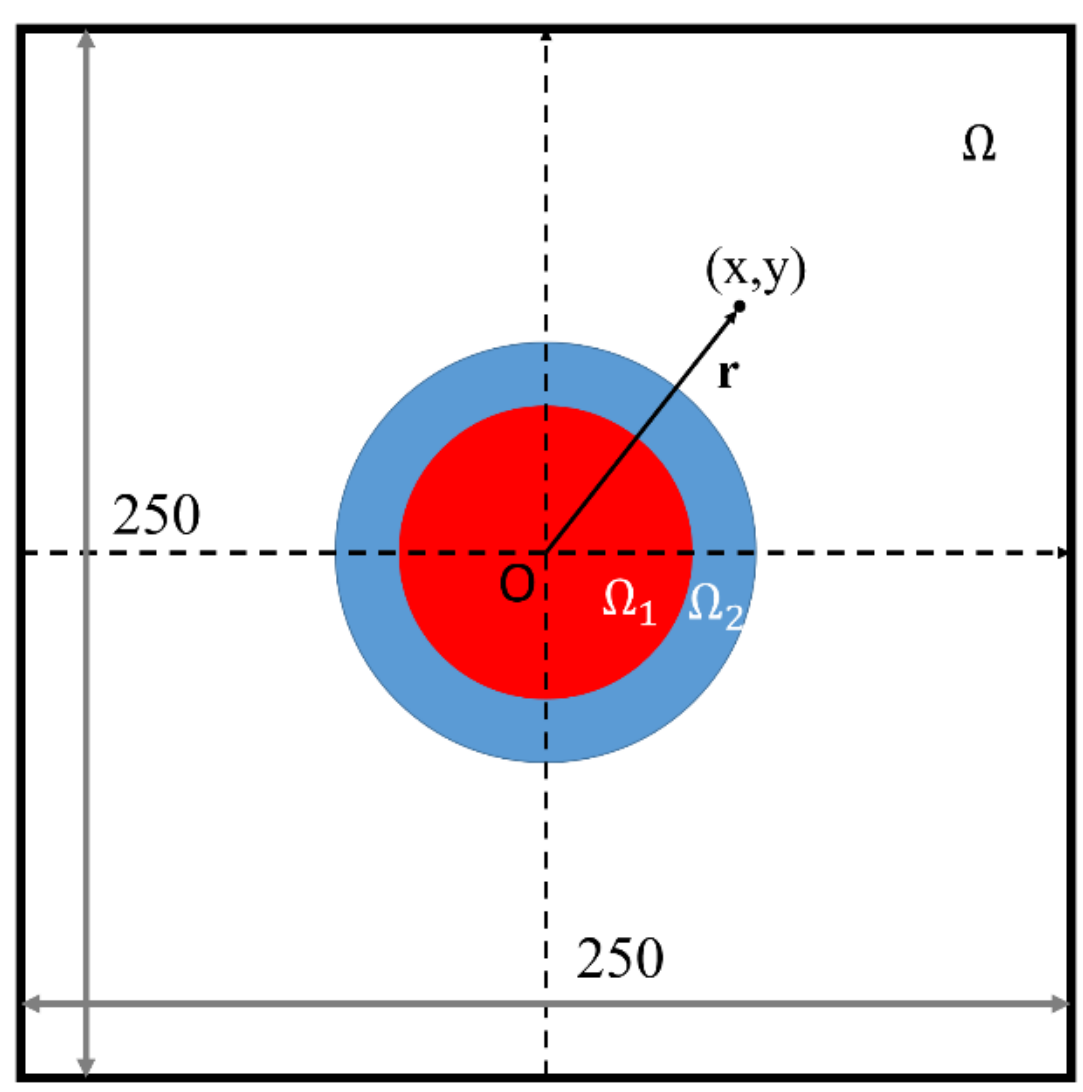

Fig. 1. Schematic of homogeneous two-dimensional computational domain, Ω measuring 250×250 pixels. The red area represents the central circular source region $\Omega_1$ ($|\mathbf{r}| \leq 70$ pixels), which is surrounded by the blue concentric annular region $\Omega_2$ ($70 < |\mathbf{r}| \leq 100$ pixels). The point 'O' at the center denotes the coordinate origin.

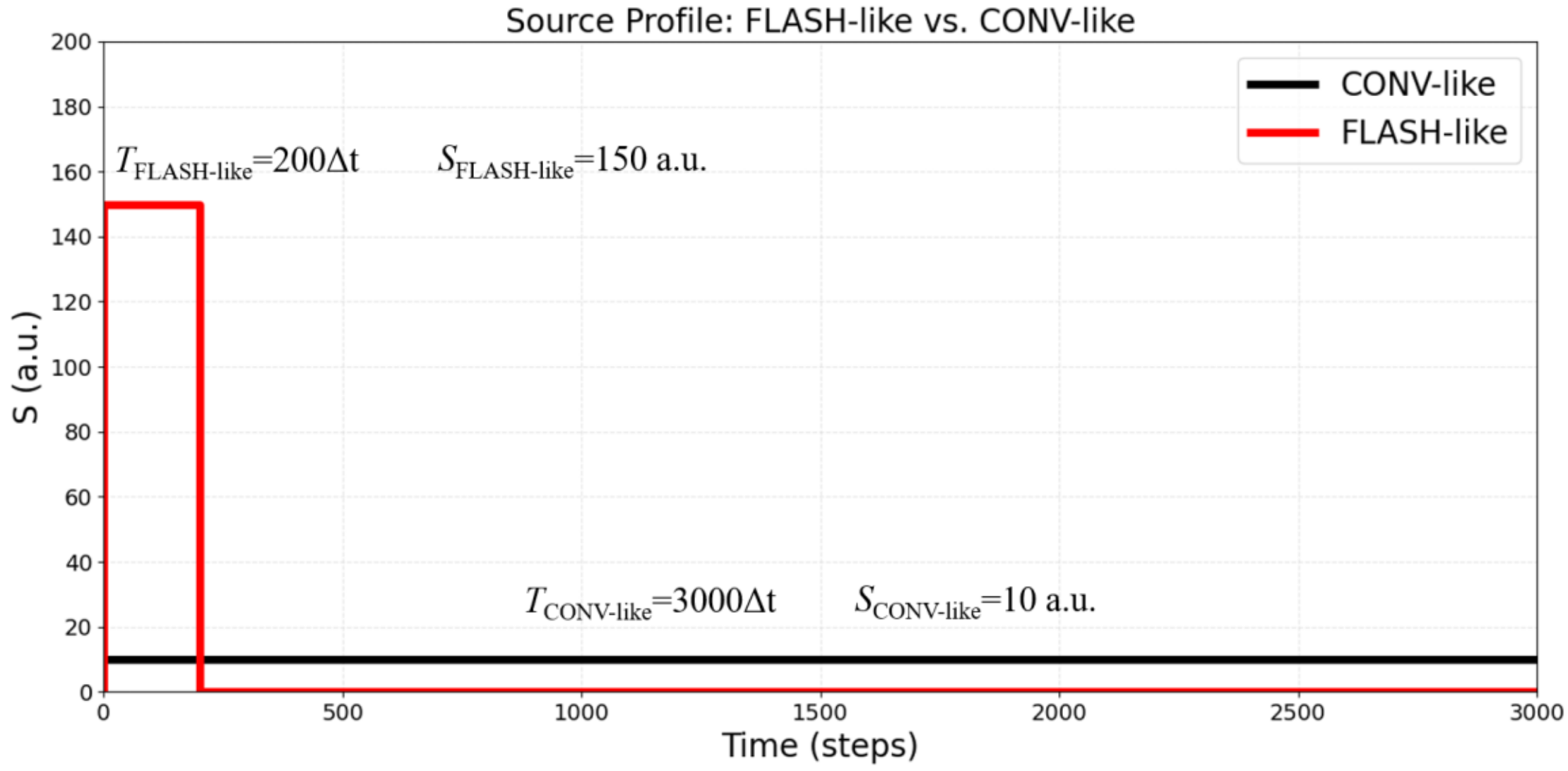

Fig. 2. CONV-like scheme delivers a continuous (0 to 3000 steps, $T_{CONV\text{-}like}=3000\cdot\Delta t$) low-intensity ($S_{CONV\text{-}like}=10$ a.u.) energy source. The FLASH-like scheme applies a rapid (0 to 200 steps, $T_{FLASH\text{-}like}=200\cdot\Delta t$) high-intensity ($S_{FLASH\text{-}like}=150$ a.u.) energy burst. The area under the curve for both source profiles is identical, representing an equivalent total incident energy deposition.

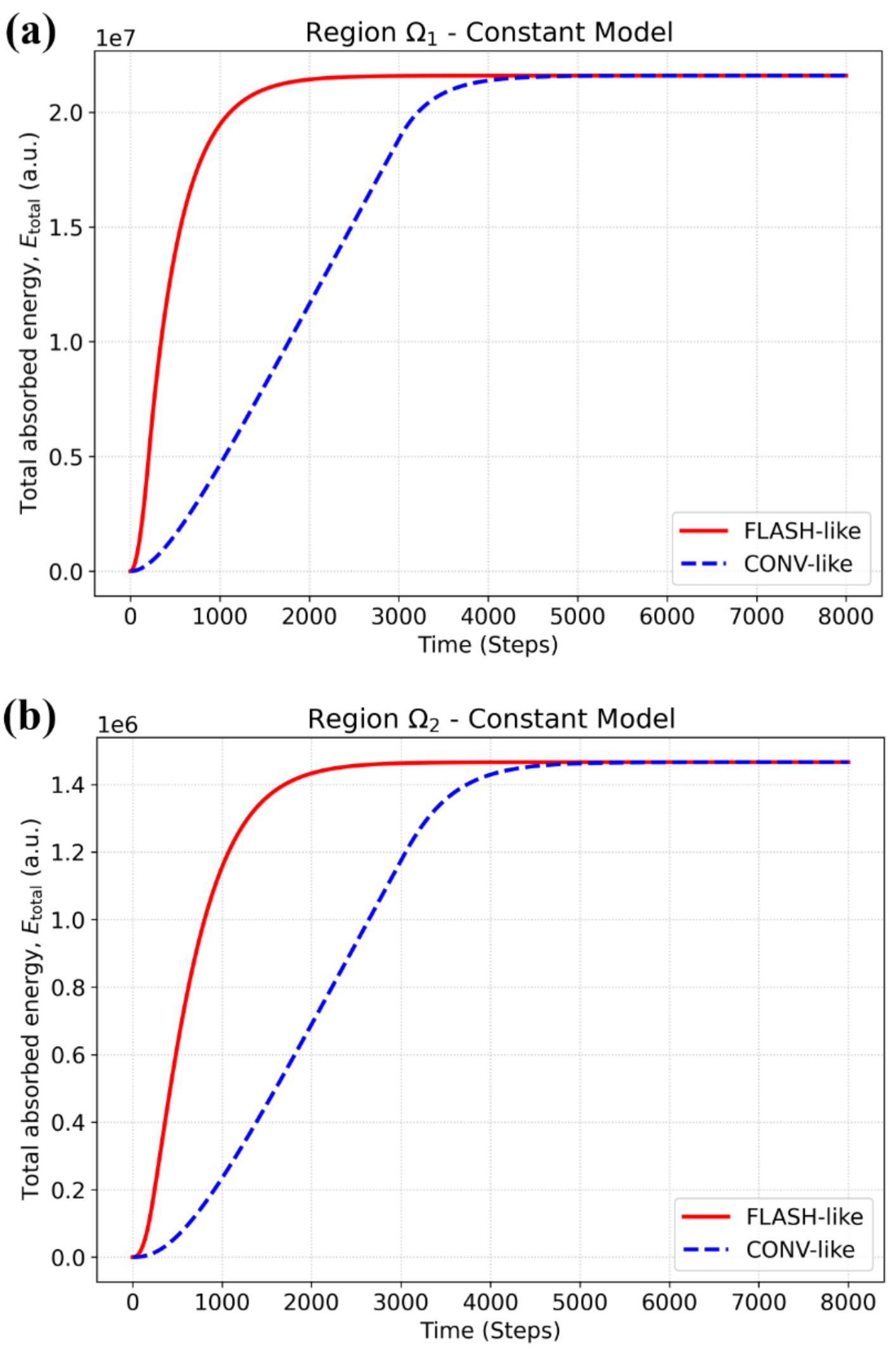

Fig. 3. Time evolution of total absorbed energy in (a) region $\Omega_1$ and (b) region $\Omega_2$ under the Constant model ($\alpha(E) = \alpha_0,\ A(E) = \alpha_0 \cdot E$). The total absorbed energies of both schemes converged to identical steady-state values at the end of the simulation, indicating that under the constant model, the final energy distribution is independent of the energy deposition schemes. These findings indicate that in the simulations, the spatial differences in energy were determined entirely by the selected nonlinear absorption model.

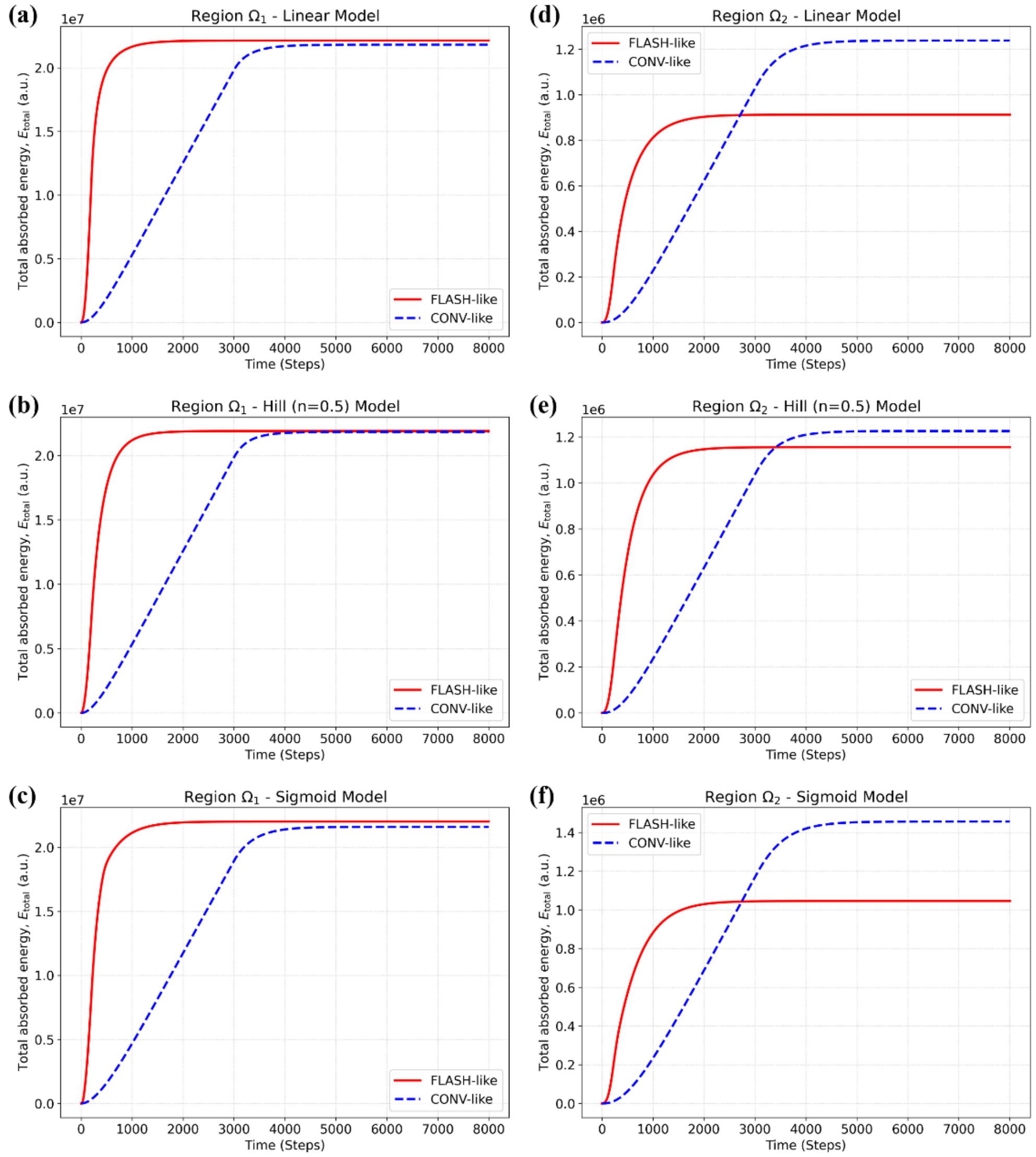

Fig. 4. Total energy absorption over time in (a-c) region $\Omega_1$ and (d-f) region $\Omega_2$ under the selected absorption models: (a, d) Linear, (b, e) Hill (n=0.5), and (c, f) Sigmoid. Under all absorption models, the FLASH-like scheme yielded lower final energy absorption in $\Omega_2$. Energy suppression was more pronounced under the Linear and Sigmoid models than under the sublinear Hill model (n=0.5).

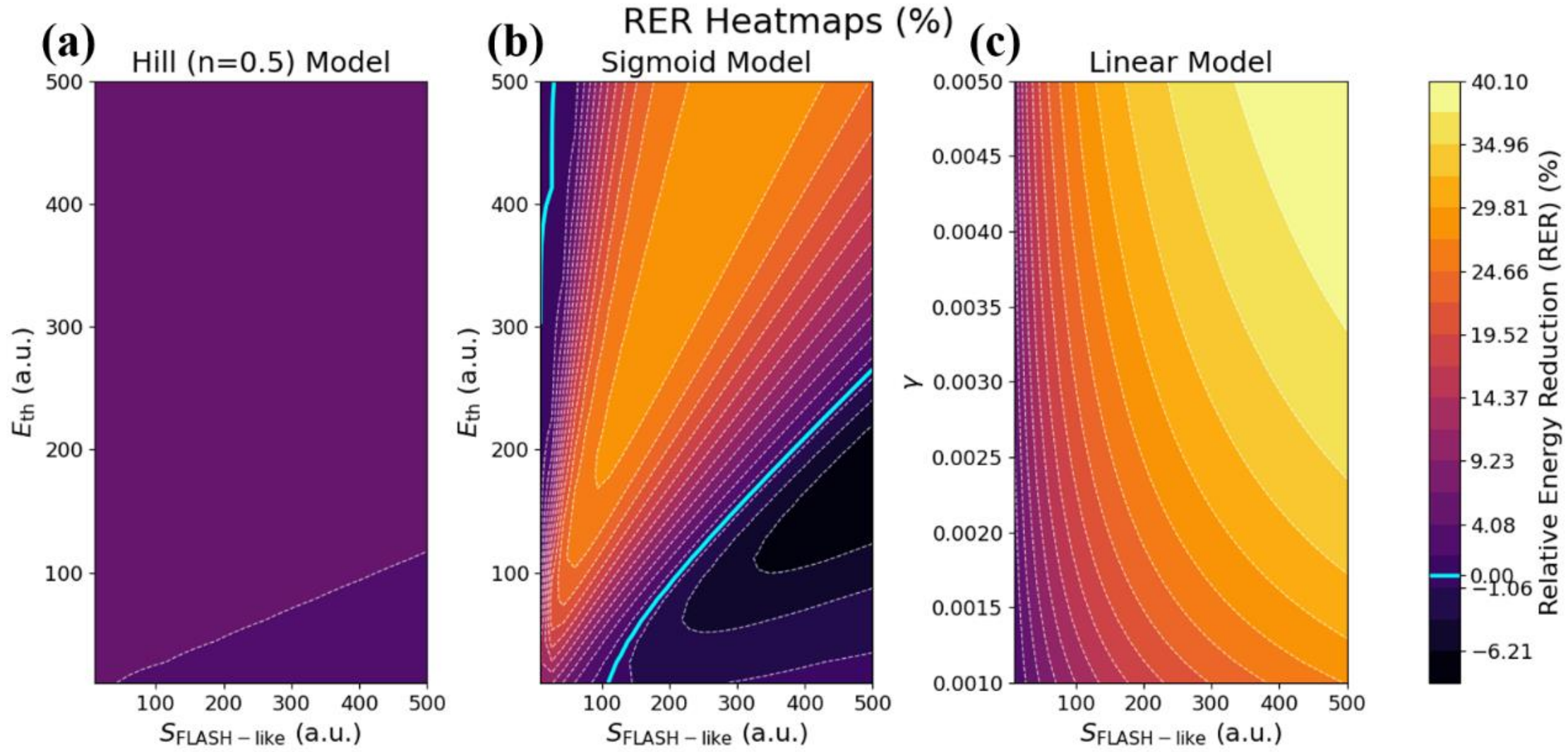

Fig. 5. Relative energy reduction (RER) heatmaps under fixed excitation durations, with $T_{\text{FLASH-like}}$ and $T_{\text{CONV-like}}$ maintained at their previously defined values, for the (a) Hill (n = 0.5), (b) Sigmoid, and (c) Linear models. FLASH-like intensity ($S_{\text{FLASH-like}}$) was varied from 10 to 500 a.u., while CONV-like intensity ($S_{\text{CONV-like}}$) was adjusted at each level to maintain the same total incident energy in both schemes.

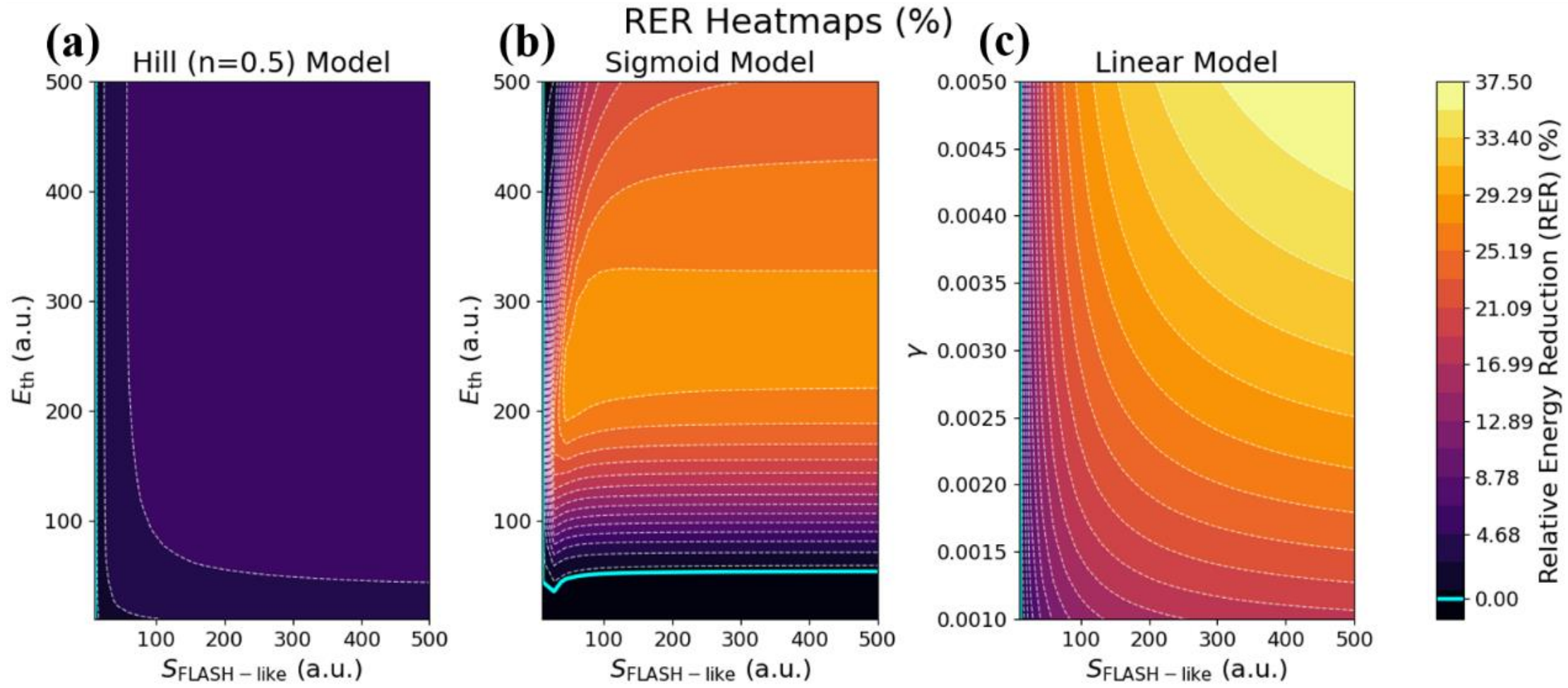

Fig. 6. Relative energy reduction (RER) heatmaps under a fixed total incident energy for the (a) Hill ($n$ = 0.5), (b) Sigmoid, and (c) Linear models. FLASH-like intensity ($S_{\text{FLASH-like}}$) was varied from 10 to 500 a.u., while the corresponding excitation duration ($T_{\text{FLASH-like}}$) was inversely scaled to maintain identical total incident energy deposition. The CONV-like intensity ($S_{\text{CONV-like}}$) and excitation duration ($T_{\text{CONV-like}}$) were held at their previously defined values.